\documentclass[%
 reprint,
 superscriptaddress,
 amsmath,amssymb,
 aps,
 pra,
]{revtex4-2}

\usepackage{xcolor}
\usepackage{hyperref}
\hypersetup{colorlinks=true,
            linkcolor=blue,
            urlcolor=blue,
            citecolor=blue}
\usepackage{graphicx}% Include figure files
\usepackage{dcolumn}% Align table columns on decimal point
\usepackage{bm}% bold math
\begin{document}

\preprint{APS/123-QED}

\title{Photo-excitation of atoms by cylindrically polarized Laguerre-Gaussian beams}% Force line breaks with \\

\author{S.~Ramakrishna}
\affiliation{Helmholtz-Institut Jena, Fr\"o{}belstieg 3, D-07743 Jena, Germany}%
\affiliation{GSI Helmholtzzentrum f\"ur Schwerionenforschung GmbH, Planckstrasse 1, D-64291 Darmstadt, Germany}
\affiliation{Theoretisch-Physikalisches Institut, Friedrich-Schiller-Universit\"at Jena, Max-Wien-Platz 1, D-07743
Jena, Germany}

\author{J.~Hofbrucker}
\affiliation{Helmholtz-Institut Jena, Fr\"o{}belstieg 3, D-07743 Jena, Germany}
\affiliation{GSI Helmholtzzentrum f\"ur Schwerionenforschung GmbH, Planckstrasse 1, D-64291 Darmstadt, Germany}

\author{S.~Fritzsche}
\affiliation{Helmholtz-Institut Jena, Fr\"o{}belstieg 3, D-07743 Jena, Germany}%
\affiliation{GSI Helmholtzzentrum f\"ur Schwerionenforschung GmbH, Planckstrasse 1, D-64291 Darmstadt, Germany}
\affiliation{Theoretisch-Physikalisches Institut, Friedrich-Schiller-Universit\"at Jena, Max-Wien-Platz 1, D-07743
Jena, Germany}

\date{\today}% It is always \today, today,

\begin{abstract}
We analyze the photo-excitation of atoms with a single valence electron by cylindrically polarized Laguerre-Gaussian beams. Theoretical analysis is performed within the framework of first-order perturbation theory and by expanding the vector potential of the Laguerre-Gaussian beam in terms of its multipole components. For cylindrically polarized Laguerre-Gaussian beams, we show that the (magnetic) sub-components of electric-quadrupole field vary significantly in the beam cross-section with beam waist and radial distance from the beam axis. We discuss the influence of varying magnetic multipole component in the beam cross-section on the sub-level population of a localized atomic target. In addition, we calculate the total excitation rate of electric quadrupole transition ($4s \;^{2}S_{1/2} \rightarrow 3d \;^{2}D_{5/2}$) in a mesoscopic target of Ca$^{+}$~ion. These calculations shows that the total rate of excitation is sensitive to the beam waist and the distance between center of the target and the beam axis. However, the excitation by cylindrically polarized Laguerre-Gaussian beam is found more efficient in driving electric quadrupole transition in the mesoscopic atomic target than the circularly polarized beams.   

\end{abstract}

\maketitle

%\tableofcontents

\section{Introduction}
 The pioneering work of Allen and coworkers~\cite{Allen1992} in 1992 showed that the twisted light beam possesses orbital angular momentum (OAM). Since then, research on twisted light (or vortex) beams has been a burgeoning research area in the scientific community~\cite{Yao,opticaloam1,opticaloam2,twistedlight}. The existence of an extra angular momentum in the vortex beam is associated with an azimuthally varying phase of the beam \cite{padgett2004light}. The presence of OAM in the twisted light beams makes it very suitable for a number of applications such as, high harmonic generation \cite{Willi} and optical tweezers \cite{opticaltweezers}. Moreover, twisted light beams have been utilized as alphabets to encode information beyond one bit per single photon \cite{twistalphabets}. This makes the Laguerre-Gaussian (LG) beams a promising candidate for high-dimensional quantum information \cite{highquantuminfo}, quantum memories \cite{quantummemory} and quantum cryptography \cite{quantumcrypto}.   

\par Atomic processes with twisted light beams, such as photo-ionization of atoms \cite{photoionization1,photoionization2,angulardistribution} and the scattering of twisted light beams by ions \cite{rayleighscatt} or electrons \cite{comptonscattering} have also gained much attention.  In addition to photo-ionization and scattering processes, the photo-excitation of atoms \cite{selection1,selection3,duan2019} by twisted light beams has attracted much interest in recent years. In a recent experiment, Schmiegelow \textit{et al.} \cite{schmiegelow2016} observed for the first time the transfer of OAM from LG beams to the bound electrons of an atom during the excitation process. This experiment investigated the electric-quadrupole transition in the Ca$^{+}$~ion positioned on the beam axis and demonstrated the suppression of AC-Stark shift in the dark centre of the LG beams. Later, a theoretical study \cite{peshkov} showed that the OAM is imprinted in the magnetic sub-level population and fluorescence following the excitation by a twisted light beam.
\par In contrast to the plane waves, twisted light beams support many different states of polarization, such as cylindrical polarization. However, photo-excitation of atoms by cylindrically polarized LG beams remains less explored. In this work, we analyze the multipole distribution of the cylindrically polarized LG beams and thus discuss its influence on the magnetic sub-level population in the target atom. Furthermore, for a target with Gaussian spatial distribution we calculate the total rate of excitation. Our calculation shows that the cylindrical polarization is more efficient than circular polarization in driving the electric quadrupole transition in the target atoms. As an example, we consider the electric-quadrupole transition $4s\;^{2}S_{1/2} \rightarrow 3d\;^{2}D_{5/2}$ in the Ca$^{+}$ ion. This particular transition has already been observed in the past experimental as well as studied in theoretical works and thus serves as an ideal case for a comparison of our theoretical work.   
\par This paper is structured as follows. In Sec. \ref{subsec: Circularly polarized LG beams} we first recall circularly polarized LG beams. We then use the knowledge of circularly polarized LG beams to understand the cylindrical polarization and to construct the vector potential of cylindrically polarized LG beams in Sec. \ref{subsec:Cylindrically polarized LG beams}. Moreover, we expand the vector potential of LG beam into its multipole components to obtain complex weight factors for a given polarization in Sec. \ref{subsec:Multipole expansion of a vector potential of the LG beams}. In the Sec. \ref{subsec:Interaction of LG beams with the target atom} we use the multipole expansion of vector potential to derive the transition amplitude for the photo-excitation of atoms by cylindrically polarized LG beams. We evaluate strength of the electric-quadrupole field in the beam cross section of a radially polarized LG beam with respect to circularly polarized LG beam in Sec. \ref{subsec:Distribution of electric-quadrupole component in the beam cross section of cylindrically polarized LG beam}.  Furthermore, in Sec. \ref{subsec:Effect of the beam waist and radial position on the multipole distribution of LG beams} we analyze distribution of the projection of electric-quadrupole field in the beam cross section with respect to the beam waist $w_{o}$ and radial distance $b$ from the beam axis of cylindrically polarized LG beams. In the first part of Sec. \ref{subsec:photoexcitation} we illustrate the photo-excitation of a well-localized target by cylindrically polarized LG beams and in the second part we calculate the total rate of excitation for electric quadrupole transition between 4s $^{2}S_{1/2}$ $\rightarrow$ 3d $^{2}D_{5/2}$ state in Ca$^{+}$~ion target with a Gaussian spatial distribution. Finally, a summary of the paper is given in the Sec. \ref{sec:summary}

\section{Theoretical background}
\subsection{Circularly polarized LG beams} 
\label{subsec: Circularly polarized LG beams}
LG beams are paraxial twisted light beams whose amplitude distribution $u(r)$ is known to satisfy the paraxial wave equation \cite{siegman1986lasers}
 \begin{equation}
        \nabla^{2}\;u\;+2\;i\;k\;\frac{\partial}{\partial z}u \; = \; 0.
    \end{equation}
This \textit{paraxial wave} approximation is valid, if the amplitude distribution $u(\bm{r})$ changes slowly with the distance $z$ and this $z$ dependence is less compared to variations of $u(\bm{r})$ in the transverse direction 
\begin{equation}
        \left|\frac{\partial^{2}u}{\partial z^{2}}\right|\ll \left|2k \; \frac{\partial u}{\partial z}\right| \; , \quad \left|\frac{\partial^{2}u}{\partial z^{2}}\right| \ll \left|\partial_{t}^{2}u\right|.
\end{equation}
In cylindrical coordinates the amplitude distribution of the LG beam is given by
 \begin{widetext}
 \begin{equation}\label{eq:amplitude}
        u(\bm{r}) = \frac{1}{w(z)} \left( \frac{\sqrt{2}r}{w(z)} \right)^{m_{\ell}} \textrm{exp}\left(- \frac{r^{2}}{w^{2}(z)} \right) L^{m_{\ell}}_{p}\left(\frac{2r^{2}}{w^{2}(z)} \right)
              \textrm{exp}\left[ im_{\ell}\phi + \frac{ikr^{2}z}{2(z^{2}+z^{2}_{R})} - i(2p+m_{\ell}+1)  \textrm{arctan}\left(\frac{z}{z_{R}}\right) \right],
 \end{equation}
where $m_{\ell}$ is the projection of the OAM upon the propagation axis, $p$ is the radial index of the LG beam, $z_{R}$ is the (so-called) Rayleigh range, $w(z)$ is the beam width of the LG beam. The beam width $w(z)$ of the LG beam varies along the propagation distance and is minimum for $z = 0$. This minimum beam width of a LG beam is known also as beam waist $w_{o}$ $\equiv w(z = 0)$. Moreover, equation (\ref{eq:amplitude}) can be used to obtain the intensity distribution $|u(r)|^{2}$ of the LG beam which exhibits a concentric ring-like structure in the beam cross section.
\par The wave amplitude of the circularly polarized LG beam in momentum space is expressed as a Fourier transformation of the amplitude distribution (\ref{eq:amplitude}). Then the vector potential of a circularly polarized LG beam in Coulomb gauge is given by (see \cite{peshkov} for a detailed derivation)
 \begin{equation} \label{eq:vectorpotential}
        \mathbf{A}^{\mathrm{(circ)}}_{m_{\ell},\lambda,p}\bm{(r)} = \int d^{2}\bm{k}_{\bot} v_{pm_{\ell}}(k_{\bot})\; e^{i(m_{\ell}+\lambda)\phi_{k}}\; \mathbf{e}_{k,\lambda} \;e^{i \mathbf{k}\cdot \mathbf{r}} ,
\end{equation}
where $v_{pm_{\ell}}(k_{\bot})\; e^{i(m_{\ell}+\lambda)\phi_{k}}$ is the momentum space wave function and $\mathbf{e}_{k,\lambda} \;e^{i \mathbf{k}\cdot \mathbf{r}}$ is the vector potential of a circularly polarized plane waves.  In the above equation momentum space wave function $v_{pm_{\ell}}(k_{\bot})$ is given by
 \begin{equation}
        v_{pm_{\ell}}(k_{\bot}) = \frac{(-i)^{m_{\ell}}}{ w_{o} 4\pi} \; e^{-k_{\bot}^{2}w^{2}_{o}/4} \; \left(\frac{k_{\bot} w_{o}}{2}\right)^{m_{\ell}}  
         \sum^{p}_{\beta=0} (-1)^{\beta} \; 2^{\beta + m_{\ell}/2} \; \left( p+m_{\ell} \atop p-\beta   \right) \;L^{m_{\ell}}_{\beta}\left( \frac{k_{\bot}^{2}w^{2}_{o}}{4} \right).
\end{equation}
\end{widetext}
As seen from equation (\ref{eq:vectorpotential}), the LG beams with circular polarization can be expressed as a coherent superposition of circularly polarized plane waves in the momentum space. The momentum vector $\bm{k}$ of these plane waves lie on the surface of a cone in the momentum space with an opening angle of $\theta_{k}$ = arctan$(k_{\bot}/k_{z})$. 

\subsection{Cylindrically polarized LG beams}
\label{subsec:Cylindrically polarized LG beams}
Cylindrically polarized LG beams can be constructed as a linear combination of two circularly polarized LG modes \cite{cylindrica}. This results in non-separable spatial and polarization modes \cite{andrews2012}. As a consequence of non-separable spatial and polarization modes, the state of polarization across the beam cross section in cylindrically polarized LG beams is spatially in-homogeneous \cite{nonuniformsop}. Beams which are linear combinations of two LG modes are known as vector beams \cite{vectorsolution} and constitute a vector solutions to the paraxial wave equation. 
\par Radial and azimuthal polarizations are two special cases of a cylindrical polarization constructed as a linear combinations of two LG modes with the projection of OAM $m_{\ell} = \pm1$ and helicity $\lambda = \mp1$ \cite{linearcombinationofcircular}. The electric field of the radially polarized LG beam always points in the radial direction and is perpendicular to the beam axis \cite{radiallypol,sabrina}. The vector potential of a radially polarized LG beam is constructed as a linear combination of vector potential of right circularly polarized $\mathbf{A}^{\mathrm{(circ)}}_{m_{\ell}=-1,\lambda=1,p}(\mathbf{r})$ and left circularly polarized $\mathbf{A}^{\mathrm{(circ)}}_{m_{\ell}=1,\lambda=-1,p}(\mathbf{r})$ LG beams and is given by
\begin{equation}
        \mathbf{A}^{\mathrm{(rad)}}_{p}(\mathbf{r}) = \frac{-i}{\sqrt{2}} \left[ \bm{A}^{\mathrm{(circ)}}_{m_{\ell}=1,\lambda =-1,p}(\mathbf{r}) + \bm{A}^{\mathrm{(circ)}}_{m_{\ell}=-1,\lambda =1,p}(\mathbf{r}) \right].
\end{equation}

\par Similarly, the electric field direction of the azimuthally polarized LG beams is always perpendicular to the radial direction \cite{azimuthal,sabrina} and its vector potential is given by
 \begin{equation}
        \mathbf{A}^{\mathrm{(azmth)}}_{p}(\mathbf{r}) = \frac{1}{\sqrt{2}} 
        \left[ \bm{A}^{\mathrm{(circ)}}_{m_{\ell}=1,\lambda=-1,p}(\mathbf{r}) - \bm{A}^{\mathrm{(circ)}}_{m_{\ell}=-1,\lambda=1,p}(\mathbf{r})   \right].
\end{equation}
Further, we use the vector potential of LG beam of a given polarization to obtain the complex weight factors using the multipole expansion in the next section.

\subsection{Multipole expansion of a vector potential of the LG beam }
\label{subsec:Multipole expansion of a vector potential of the LG beams}
A multipole expansion of the radiation field enables us generally to expand the vector potential in angular momentum basis \cite{brinkangularmomentum,rose1995elementary}. Since the intensity distribution of the LG beam is nonuniform in the beam cross section \cite{molina2007twisted}, we perform multipole expansion of the vector potential of a LG beam at a radial distance $b$ from the beam axis. That is, when the z axis is translated by a vector $\bm{b} = b\bm{e_{x}}$ from the beam axis. Then the multipole expansion of such a vector potential at a position b from the beam axis, for circularly polarized LG beam is given by
\begin{equation}
        \small{\mathbf{A}^{\mathrm{(circ)}}_{m_{\ell},\lambda,p}(\bm{r};b,w_{o} ) \; = \; \sum_{L,M,\Lambda} W^{\mathrm{(circ)}}_{m_{\ell},\lambda,p}(L,M,\Lambda; b,w_{o}) \; \mathbf{a}^{\Lambda}_{L,M}\mathbf{(r)}},
\end{equation}
where $L$ and $M$ are the eigenvalues of the TAM and the projection of TAM operators, respectively. The $W^{\mathrm{(circ)}}_{m_{\ell},\lambda,p}(L,M,\Lambda;b,w_{o})$ is the complex weight factor of the expansion which depends on the radial distance from the beam axis $b$, the beam waist $w_{o}$, projection of orbital angular momentum $m_{\ell}$, the radial index $p$ and the helicity $\lambda$. The multipole expansion expresses the vector potential of the LG beam as a linear combination of electric ($\Lambda = 1$) and magnetic ($\Lambda = 0$) multipole components $\mathbf{a}^{\Lambda}_{L,M}\mathbf{(r)}$. Mathematically, $\mathbf{a}^{\Lambda}_{L,M}\mathbf{(r)}$ are expressed in terms of vector spherical harmonics of rank $L$ \cite{reftovectorspherical,johnson2007atomic}. For circularly polarized LG beams, the complex weight factors is given by
\begin{widetext}
\begin{align}
        W^{\mathrm{(circ)}}_{m_{\ell},\lambda,p}(L,M,\Lambda;b,w_{o}) &= \sum_{L,M,\Lambda} \;\sum^{p}_{\beta=0} \;(i \lambda)^{\Lambda}\;(-1)^{\beta} \; 2^{\beta +\frac{m_{\ell}}{2}} \; \left( p+m_{\ell} \atop p-\beta \right) \; \frac{(-i)^{m_{\ell}}w_{o}}{2\pi}\; 
         (i)^{L+m_{\ell}+\lambda-M} \; (2 L + 1)^{1/2} \\\nonumber 
         & \times \;\int_{0}^{\infty} k_{\bot} dk_{\bot}\; d^{L}_{M,\lambda}(\theta_{k})\; e^{-\frac{k_{\bot}^{2}w_{o}^{2}}{4}} \left( \frac{k_{\bot} w_{o}}{2} \right)^{m_{\ell}} \; L^{m_{\ell}}_{\beta}\left( \frac{(k_{\bot} w_{o})^{2}}{4} \right) \; J_{m_{\ell}+\lambda-M}(k_{\bot} b).
\end{align}
\vspace{0.28in}
Using these, we define the complex weight factors for the radially polarized LG beam as 
\begin{align}
    W^{\mathrm{(rad)}}_{p}(L,M,\Lambda;b,w_{o}) = \frac{-i}{\sqrt{2}}\left[ W^{\mathrm{(circ)}}_{m_{\ell}= 1,\lambda = -1,p}(L,M,\Lambda;b,w_{o}) + W^{\mathrm{(circ)}}_{m_{\ell} = -1,\lambda = 1,p}(L,M,\Lambda;b,w_{o}) \right]
\end{align}
\vspace{0.28in}
and for the azimuthally polarized LG  beam as
\begin{align}
    W^{\mathrm{(azmth)}}_{p}(L,M,\Lambda;b,w_{o}) = \frac{1}{\sqrt{2}}\left[ W^{\mathrm{(circ)}}_{m_{\ell}= 1,\lambda = -1,p}(L,M,\Lambda;b,w_{o}) - W^{\mathrm{(circ)}}_{m_{\ell} = -1,\lambda = 1,p}(L,M,\Lambda;b,w_{o}) \right].
\end{align}

With the help of complex weight factors, we can study strength of the individual multipole components of the radiation field. Since these complex weight factors depend on the radial distance $b$ and the beam waist $w_{o}$, we can control the multipole distribution of the LG beam by carefully choosing $b$ and $w_{o}$.
\end{widetext}
\vspace{1in}
\subsection{Interaction of LG beams with the target atom}\label{subsec:Interaction of LG beams with the target atom}
The transition between initial $|\alpha_{i}J_{i}M_{i}\rangle$ and final $| \alpha_{f}J_{f}M_{f}\rangle$ bound states of an atom is given by the transition amplitude $M_{fi}$
 \begin{equation}\label{eq:transition}
        M_{fi} \; = \; \langle \alpha_{f}J_{f}M_{f}| \bm{\alpha}\cdot\mathbf{A}(\mathbf{r}) |\alpha_{i}J_{i}M_{i}\rangle,
\end{equation}
where $\mathbf{A}(r)$ is the vector potential of either radially polarized or azimuthally polarized LG beams, $\bm{\alpha}$ is the Dirac matrix and the atomic initial (final) states are characterized by TAM $J_{i}$ ($J_{f}$), TAM projection $M_{i}$ ($M_{f}$) quantum numbers and $\alpha$ refers to all additional quantum numbers, respectively.  

\par We substitute the multipole expansion of vector potential of LG beams into equation (\ref{eq:transition}) to obtain the transition amplitude
\begin{align}
        M_{fi}(b,w_{o})  &= \sum_{L,M,\Lambda} W(L,M,\Lambda;b,w_{o}) \\ \nonumber 
        & \times \langle \alpha_{f}J_{f}M_{f}|\bm{\alpha} \cdot\mathbf{a}^{\Lambda}_{L,M}(\mathbf{r}) | \alpha_{i}J_{i}M_{i} \rangle,
\end{align}
where $W(L,M,\Lambda;b,w_{o})$ is the complex weight factor of either a radially polarized or azimuthally polarized LG beam. The rest of the transition amplitude equation is solved using the Wigner-Eckart theorem \cite{brinkangularmomentum,rose1995elementary}, which gives the transition amplitude 
 \begin{align}\label{eq:tamplitude}
      M_{fi}(b,w_{o}) &= \sum_{L,M,\Lambda} W(L,M,\Lambda;b,w_{o}) \langle J_{i}M_{i},LM| J_{f}M_{f}\rangle \\ \nonumber 
     &\times  \langle \alpha_{f}J_{f}||\bm{\alpha}\cdot \mathbf{a}^{\Lambda}_{L}(\mathbf{r})||\alpha_{i}J_{i}\rangle,
\end{align}
as the product of geometrical and atomic factors. The complex weight factor $W(L,M,\Lambda;b,w_{o})$ and the Clebsch-Gordan (CG) coefficient represents the geometrical and the reduced matrix elements describe the atomic properties which influence the atomic excitation process.  It is clear from the above equation (\ref{eq:tamplitude}) that the transition amplitude depends on the complex weight factors $W(L,M,\Lambda;b,w_{o})$, which together with the CG coefficients and reduced matrix elements determine the amplitudes of the individual transitions.
\par The transition amplitude of the atomic transition between two bound states must satisfy the set of following rules known as \textit{selection rules} given by
\begin{equation}
    M_{i} + M = M_{f}
\end{equation}
\begin{equation}
     |J_{f}-J_{i}| \leq L \leq |J_{f}+J_{i}| 
\end{equation}
\begin{equation}
    \pi_{i} \pi_{f} = (-1)^{L+p+1}
\end{equation}
here $\pi_{i} \pi_{f}$ are the parity of the initial and final atomic states. 

\begin{figure}
    \centering
    \includegraphics[width=.48\textwidth]{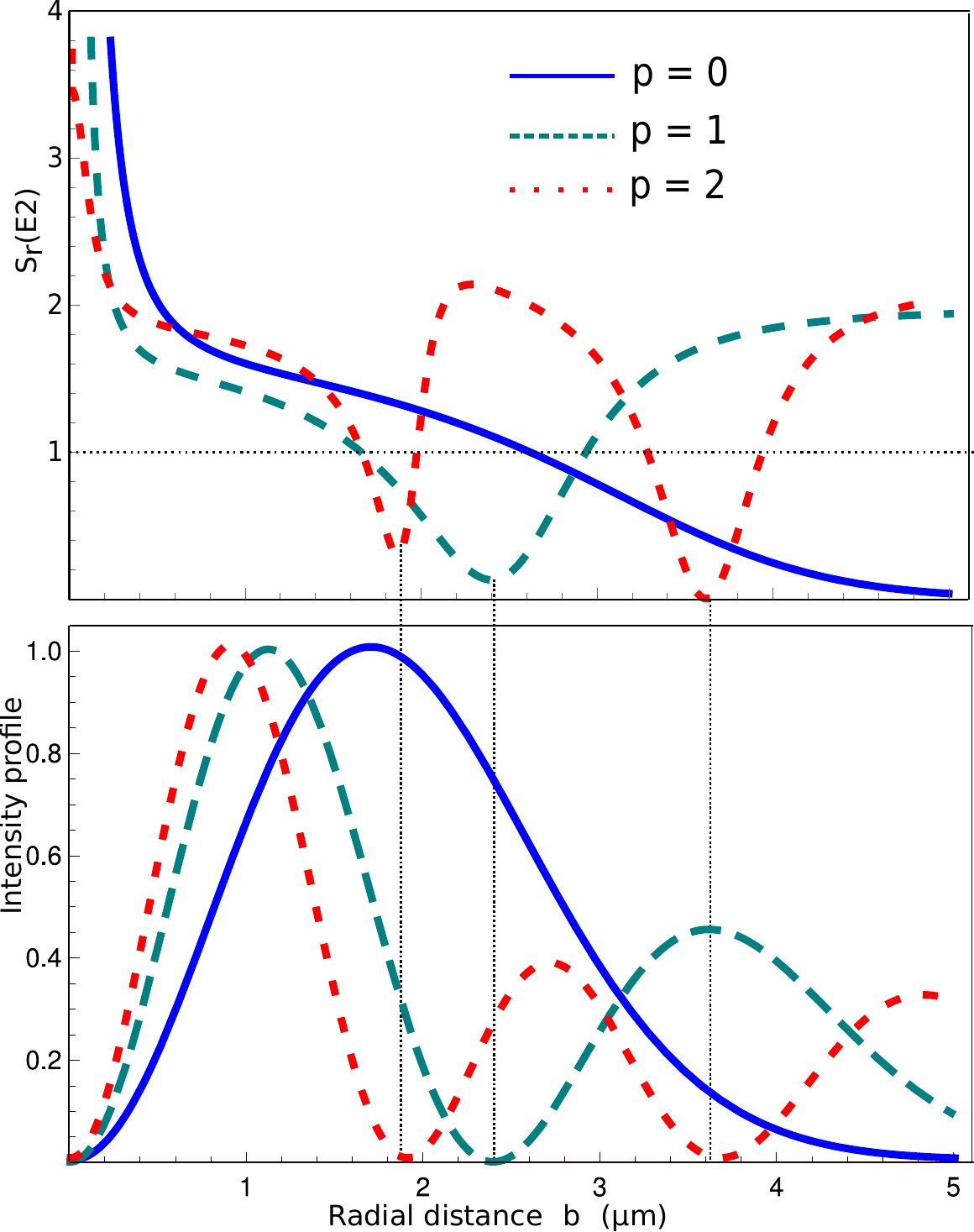}
    \caption{The relative strength of the electric-quadrupole field $S_{r}(E2)$ is plotted against the radial distance $b$ from the beam axis of the LG beam. The top plot shows variation of $S_{r}(E2)$ for three different radial indices $p = 0, 1,2$. The plot describes variation of strength of the electric-quadrupole field of radial polarization with respect to circular polarization in the beam cross section of the LG beam with beam waist of $w_{o} = 2.7$ $\mu m$. The bottom plot describes the intensity profile of the LG beams with radial index $p = 0, 1, 2$} 
    \label{fig:ratio}
\end{figure}
\section{Results and Discussion}
\subsection{Distribution of electric-quadrupole field in the beam cross section of cylindrically polarized LG beam}
\label{subsec:Distribution of electric-quadrupole component in the beam cross section of cylindrically polarized LG beam}

We now use the complex weight factors $W_{m_{\ell},\lambda,p}(L,M,\Lambda; b,w_{o})$ to analyze the distribution of the electric quadrupole field in the beam cross-section of cylindrically polarized LG beams. The strength of a multipole component $L$, in particular electric-quadrupole component (L = 2), in the beam cross section is given by~$\sum_{M=-L}^{+L}|W_{m_{\ell},\lambda,p}(L=2,M,\Lambda=1;b,w_{o})|^{2}$ for a corresponding polarization of the LG beam. We define relative strength of the electric-quadrupole~field
\begin{equation}
    S_{r}(E2) = \frac{\sum_{M=-L}^{+L}|W^{\mathrm{(circ)}}_{m_{\ell},\lambda,p}(L=2,M,\Lambda=1;b,w_{o})|^{2}}{\sum_{M=-L}^{+L}|W^{\mathrm{(rad)}}_{p}(L=2,M,\Lambda=1;b,w_{o})|^{2}}.\label{eq:WrE2}
\end{equation}
as a ratio of strength of the electric-quadrupole field between circularly polarized and radially polarized LG beams.

\par In Fig.~\ref{fig:ratio}, we plot $S_{r}(E2)$ against the radial distance $b$ from the beam axis of a LG beam for different radial index $p$ values in the top panel and the corresponding intensity profile of LG beams in bottom panel. For the radial index $p = 0$, we observe the value of $S_{r}(E2)$ to be greater than one near the beam axis ($b \approx 0$) and decreases rapidly to less than one as radial distance $b$ increases. The behaviour of the ratio $S_{r}(E2)$ indicates that the strength of the electric-quadrupole field is suppressed near the beam axis of cylindrically polarized LG beam. However, for large $b$ values the ratio $S_{r}(E2)$ is less than one, indicating a strong electric-quadrupole field in the beam cross section of cylindrically polarized LG beam. The radial index $p$ of the LG beam modifies strength of the electric-quadrupole field in the beam cross section as described by the Fig.~\ref{fig:ratio} for $p = 1, 2$. Similar to $p = 0$ case, the ratio $S_{r}(E2)$ is greater than one near the beam axis, indicating the suppression of an electric-quadrupole field near the beam axis for cylindrically polarized LG beam. For large $b$ values, we observe the ratio $S_{r}(E2)$ to be lesser than one near the dark region in the beam cross section of the LG beam. The vertical lines in the Fig.~\ref{fig:ratio} is used to denote the corresponding dark region in the beam cross section of LG beam with the help of intensity profile. In contrast to $p = 0$ case, cylindrically polarized LG beams possess a dominant electric-quadrupole field with respect to circularly polarized LG beams only near the dark region in the off-axis region.

\par The electric-quadrupole component can be associated with the electric field gradient of the light beam and is responsible for driving the electric-quadrupole transition in the target atoms. Therefore, the region in the Fig.~\ref{fig:ratio} with value of $S_{r}(E2)$ less than one describes a strong electric field gradient region in the beam cross section of cylindrically polarized LG beams. This suggests that, if we were to place an atom in such a region the electric-quadrupole transition would be more efficiently driven by a cylindrical polarization over a circularly polarized LG beam.

\begin{figure}
    \centering
    \includegraphics[width=0.48\textwidth]{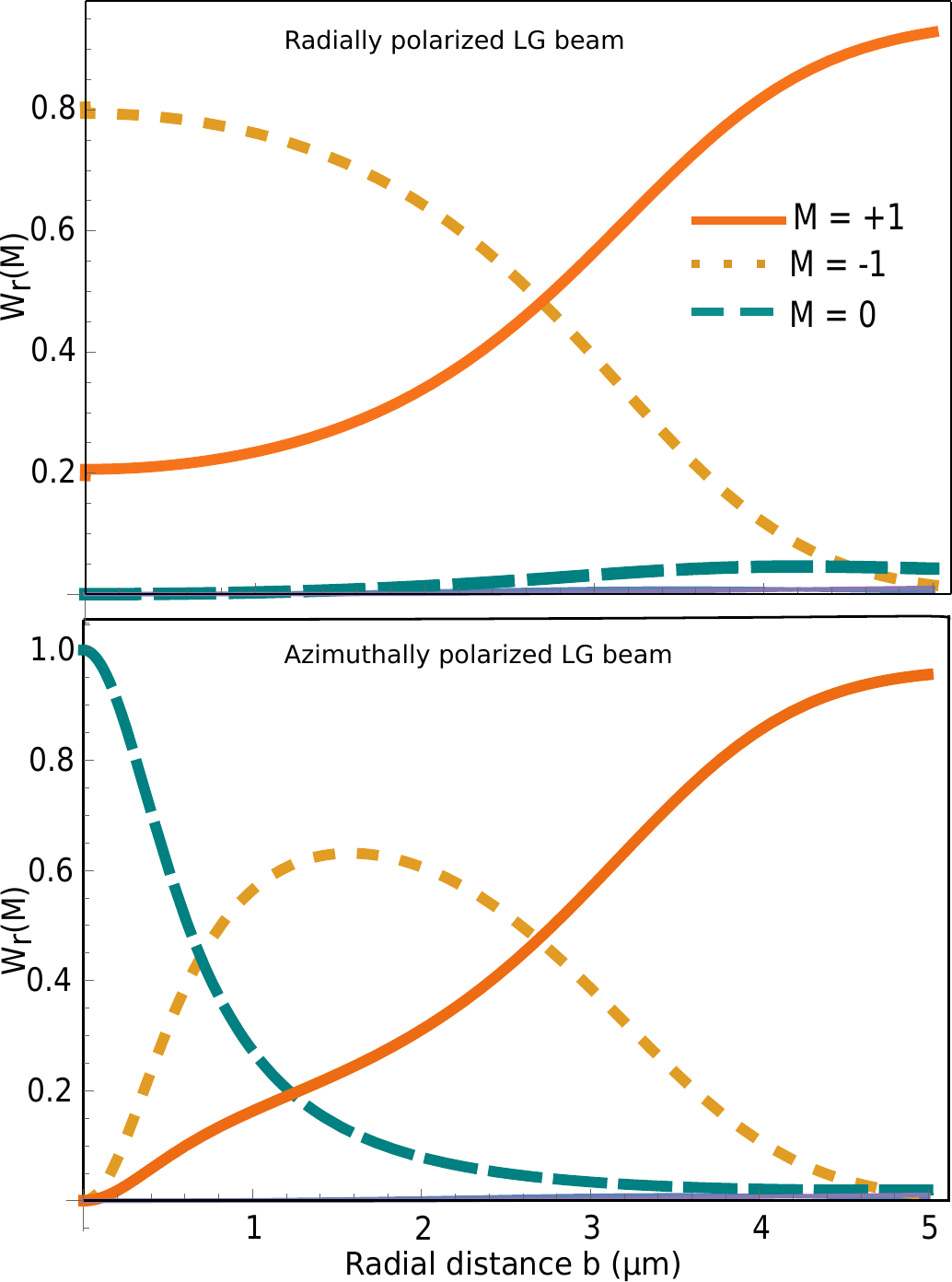}
    \caption{The relative weight of the projection of quadrupole component $W_{r}(M)$ is plotted against the radial distance $b$ from the beam axis for radially polarized (top) and azimuthally polarized LG beam (bottom). In the above plots radial index $p = 0$ and the beam waist is fixed to $w_{o} = 2.7$ $\mu m$.}
    \label{fig:cylrad}
\end{figure}
\subsection{Influence of the beam waist and radial position on the projection of TAM of LG beams}\label{subsec:Effect of the beam waist and radial position on the multipole distribution of LG beams}
We shall now discuss strength of the projection of multipole component $M$ in the beam cross section of the LG beam using the complex weight factors. As seen from the complex weight factors, the multipole distribution of the LG beams varies with the beam waist $w_{o}$ and the radial distance $b$ from the beam axis. To understand the variation of the projection of TAM across the beam cross section, we define the relative weight of projection of the electric-quadrupole field as
\begin{equation}
    W_{r}(M) = \frac{|W_{p}(L=2,M,\Lambda =1;b,w_{o})|^{2}}{\sum_{M=-L}^{+L}|W_{p}(L=2,M,\Lambda=1;b,w_{o})|^{2}}
\end{equation}
where $|W_{p}(L=2,M,\Lambda=1;b,w_{o})|^{2}$ denotes the modulus squared of the complex weight factor of a radially or azimuthally polarized LG beam. 
\par In the Fig. \ref{fig:cylrad}, we plot $W_{r}(M)$ against the radial distance from the beam axis $b$ while keeping the beam waist fixed at 2.7 $\mu m$ \cite{schmiegelow2016}. For radially polarized LG beams, on the beam axis we observe M $= -1$ component to be dominant while M = $0$ component is dominant for azimuthally polarized LG beam. For large values of $b$, we notice M $= +1$ component to be dominant for both radially and azimuthally polarized LG beams. 

\par Similarly in the Fig. \ref{fig:cylwaist}, we plot $W_{r}(M)$ against the beam waist $w_{o}$ of the cylindrically polarized LG beam for a fixed $b$ value. Since we consider a paraxial LG beam, the minimum beam waist is given by the diffraction limit
\begin{equation}
    w_{o}^{(min)} = \frac{\lambda}{\sqrt{2}\pi}
\end{equation}
where $\lambda$ = 729 nm corresponds to the wavelength of the LG beam used for driving transition between 4s and 3d states of Ca$^{+}$ ion \cite{schmiegelow2016}. For this beam waist $w_{o}$ = 0.16 $\mu m$, we observe M $= +1$ component to be dominant in both radial and azimuthal polarization. For larger beam waist~$w_{o}$, M $= -1$ and M $= 0$ components dominates in radially and azimuthally polarized LG beams respectively.

\par Thus, by carefully selecting the beam waist $w_{o}$ and radial distance $b$ we can control the relative strength of $M$ in the beam cross section of LG beam of either radial or azimuthal polarization.

\subsection{Photo-excitation of Ca$^{+}$ ion by LG beams}\label{subsec:photoexcitation}
\subsubsection{Single ion target}\label{subsubsec:localized}
\begin{figure}
    \centering
    \includegraphics[width=0.48\textwidth]{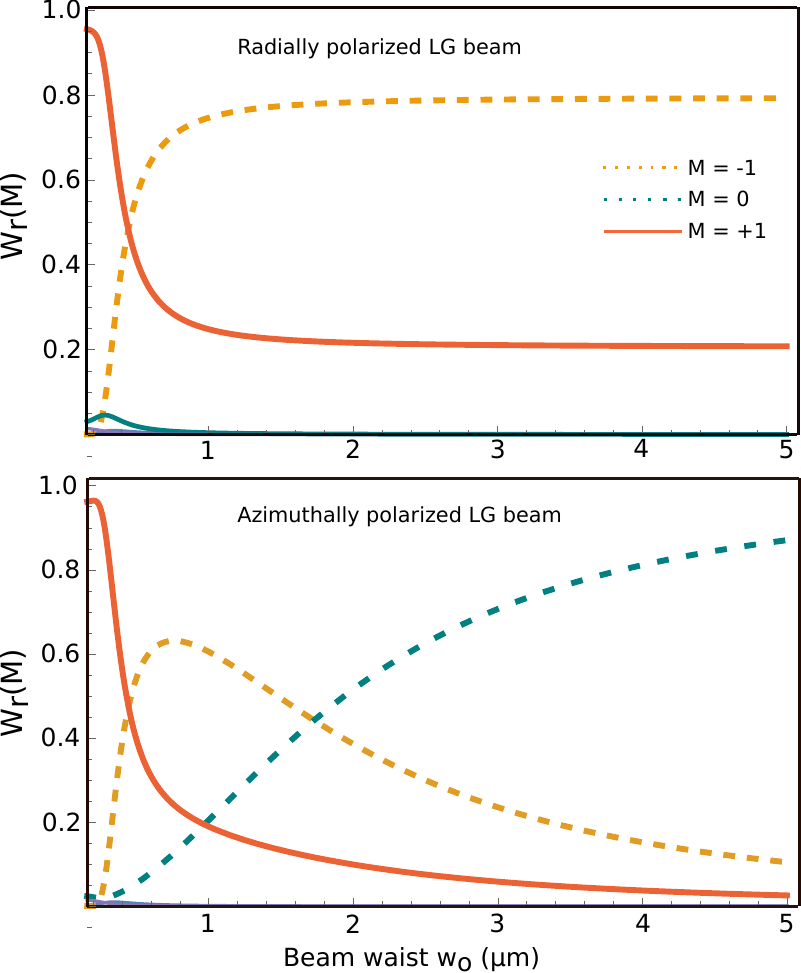}
    \caption{The relative weight of the projection of quadrupole component $W_{r}(M)$ is plotted against the beam waist $w_{o}$ for radially polarized (top) and azimuthally polarized LG beam (bottom). In the above plots radial index $p = 0$ and the radial distance is fixed to $b = 0.5$ $\mu m$.}
    \label{fig:cylwaist}
\end{figure}

We now illustrate the photo-excitation of a well-localized target by a cylindrically polarized Laguerre-Gaussian beam of beam waist $w_{o}$ and radial index $p$ with the help of Figs.~\ref{fig:cylrad}, \ref{fig:cylwaist} and the selection rule $M_{i} + M = M_{f}$. For example, we consider the electric quadrupole transition between $|4S_{1/2},M_{i} = -1/2\rangle$ and $|3D_{5/2},M_{f} =~\pm 5/2\rangle$ states in single Ca$^{+}$~ion driven by cylindrically polarized LG beams of wavelength $729$ nm \cite{schmiegelow2016}.

\par For a single Ca$^{+}$ ion positioned along the beam axis, radially polarized LG beam drives transition between $|4S_{1/2}, M_{i}=-1/2\rangle$ and the $|3D_{5/2}, M_{f}= -3/2\rangle$ magnetic sub-state transferring $M = -1$ component, as shown in the Fig.~\ref{fig:cylrad}. As the radial distance $b$ between the target Ca$^{+}$ ion and the beam axis is increased, we observe transition between $|4S_{1/2}, M_{i}=-1/2\rangle$ and the $|3D_{5/2}, M_{f}= 1/2\rangle$ magnetic sub-state absorbing $M = +1$ component. For the target Ca$^{+}$~ion positioned in the off-axis region, say, $b = 0.5$ $\mu m$ interacting with a radially polarized LG beam of minimum beam waist $w_{o} = 0.16$~$\mu m$ we observe transition between $|4S_{1/2}, M_{i} = -1/2\rangle$ and $|3D_{5/2}, M_{f}= +1/2\rangle$ magnetic sub-state absorbing $M = +1$ component as shown in the Fig.~\ref{fig:cylwaist}. But as we increase the beam waist $w_{o}$, the strength of $M = +1$ component in the beam cross section of radially polarized LG beam decreases rapidly and the strength of $M = -1$ component increases. This results in the transition between $|4S_{1/2}, M_{i} = -1/2\rangle$ and $|3D_{5/2}, M_{f}= -3/2\rangle$ magnetic sub-states.
\par Similar to the above discussion, we can explain the photo-excitation of a well localized Ca$^{+}$ ion and the corresponding atomic transitions driven by azimuthally polarized LG beam.

\subsubsection{Macroscopic target}
The well-localized point-like target considered in Sec.~\ref{subsubsec:localized} is an idialistic model and is difficult to realize in the experimental set-ups. In experiments, even a single ion target trapped in Paul traps has a thermal spatial spread \cite{schmiegelow2016}. Hence, we now consider the atomic target to have a gaussian spatial distribution as observed in the experiments \cite{schmiegelow2016,Afanasev_2018} and which is given by
\begin{equation}
    f(\bm{b}) = \frac{1}{2\pi \sigma^{2}} \; exp\left(- \frac{(\bm{b}-\bm{b_{o}})^{2}}{2\;\sigma^{2}}\right) \label{eq:gDist}
    \end{equation}
where $\sigma$ is the width of the target and vector $\bm{b_{o}}~=~b_{o}\bm{e_{x}}~\equiv b_{o,x}$ is the distance from the beam axis to the center of the target. For an atomic target with a Gaussian spatial distribution interacting with the Laguerre-Gaussian beams we define total rate of excitation as 
\begin{equation}
    W^{LG}_{fi} = \frac{2 \pi}{\alpha^{2}} \sum_{M_{f}} \int d^{2}\bm{b} f(\bm{b}) |M_{fi}(b,w_{o})|^{2}.\label{eq:meanrate}  
\end{equation}
Since we consider a target prepared in a specific magnetic state $M_{i}$, we do not sum and average over all initial states. As seen from the equations (\ref{eq:gDist}), (\ref{eq:meanrate}) the rate is proportional to the modulus squared of the transition amplitude and the Gaussian distribution of the target. The full many-electron transition amplitude for the electric quadrupole transition ($4s \;^{2}S_{1/2} \rightarrow 3d \;^{2}D_{5/2}$) was calculated using the JAC \cite{JAC} package (and GRASP \cite{GRASP}). 

\begin{figure}
    \centering
    \includegraphics[width=0.48\textwidth]{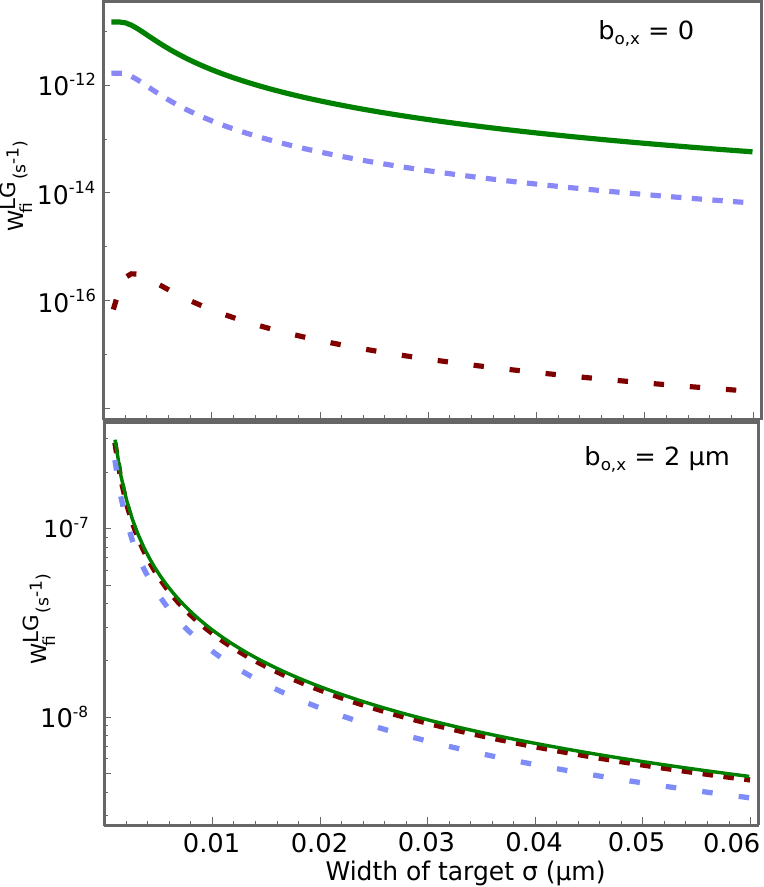}
    \caption{ Log plot of total rate of excitation for electric quadrupole transition between 4s $^{2}S_{1/2}$ $\rightarrow$ 3d $^{2}D_{5/2}$ levels of Ca$^{+}$ ion driven by LG beam is plotted as a function of width of the target $\sigma$ for azimuthal (green solid line), radial (brown dashed line) and circular polarization (blue dotted line). In the top figure the target is placed on the beam axis $b_{o,x} = 0$ and in the bottom figure the target is displaced from the beam axis by $b_{o,x}$ = 2 $\mu m$. In both the plots, radial index $p$ and beam waist $w_{o}$ is kept fixed at 0 and $2.7 \mu m$ respectively. }
    \label{fig:rate1}
\end{figure}
\par In Fig.~\ref{fig:rate1} we plot the total rate of excitation for electric quadrupole transition as a function of width of the target for the beam waist of cylindrically polarized LG beam of $w_{o}$ = 2.7 $\mu m$. In addition, we show the total rate of excitation for circularly polarized LG beams as a comparison. The width of the target $\sigma$ is varied from 0 to a maximum of 0.06 $\mu m$ as observed in the experiments. For the target placed on the beam axis, we observe the azimuthally polarized LG beam to be more efficient in driving the electric quadrupole transition for a target with small as well as large width $\sigma$. In the bottom of Fig.~\ref{fig:rate1}, we consider the target to be placed near the maximum intensity region in the beam cross-section that is, when x component of $\bm{b_{o}}$ ($b_{o,x}$) = 2 $\mu m$. For such a case, we observe both circular and cylindrical polarization to have the same $W_{fi}^{LG}$ for small width of the target. But as we increase the width of the target, we observe $W_{fi}^{LG}$ to decrease rapidly for both cylindrical as well as circularly polarized LG beam. However, for larger width of the target we notice that $W_{fi}^{LG}$ to be more for cylindrical polarization than circular polarization. As shown in the Fig.~\ref{fig:ratio} the strength of the electric quadrupole field for the cylindrically polarized LG beam increases as we move away from the beam axis, hence we obtain a maximum $W_{fi}^{LG}$ than circular polarization. 

\par
\begin{figure}
    \centering
    \includegraphics[width=0.48\textwidth]{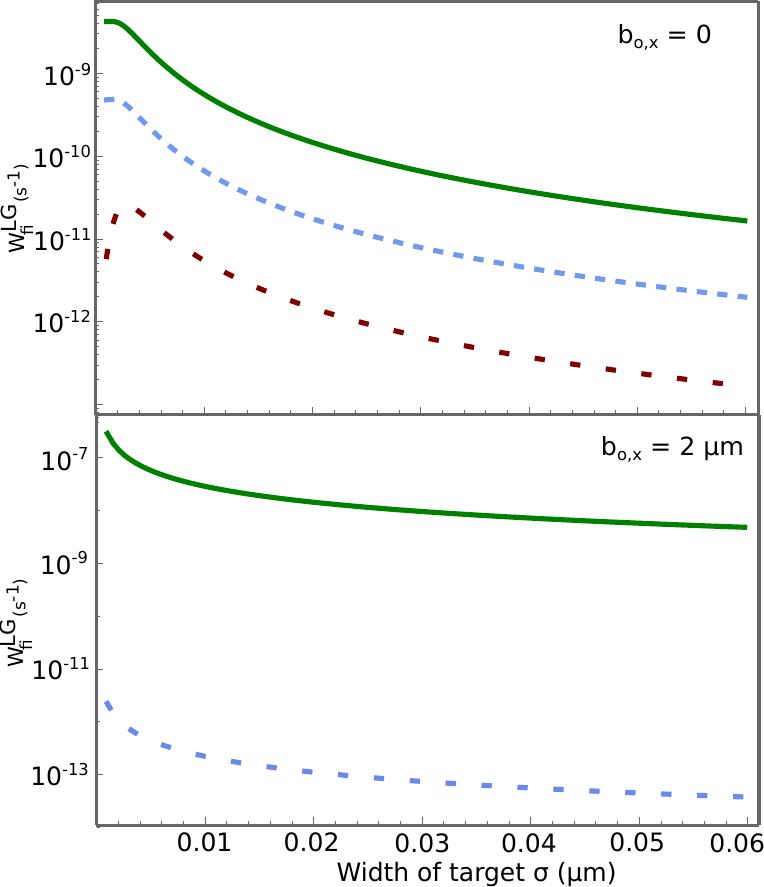}
    \caption{ Log plot of total rate of excitation for electric quadrupole transition between 4s $^{2}S_{1/2}$ $\rightarrow$ 3d $^{2}D_{5/2}$ levels of Ca$^{+}$ ion driven by LG beam is plotted as a function of width of the target $\sigma$ for both azimuthal (green solid line), radial (brown dashed line) and circular (blue dotted line) polarization. In the top figure the target is placed on the beam axis $b_{o,x} = 0$ and in the bottom figure the target is displaced from the beam axis by $b_{o,x}$ = 2 $\mu m$. In both the plots, radial index $p$ and beam waist $w_{o}$ is kept fixed at 0 and $ 0.16 \mu m$ respectively.}
    \label{fig:rate2}
\end{figure}

 In Fig. \ref{fig:rate2} the variation of the total rate of excitation is plotted as a function of width of the target placed in the center of the beam and displaced from the beam axis for the minimum possible beam waist $w_{o}$~=~0.16 $\mu m$. Similar to the case for beam waist $w_{o}$ = 2.7 $\mu m$, we observe the azimuthal polarization to have maximum total rate of excitation when compared to circular and radial polarization as shown in the top figure. In contrast to the beam waist $w_{o}$ = 2.7 $\mu m$, we observe higher total rate of excitation for radial polarization when we consider $w_{o}$~=~0.16 $\mu m$. For the target placed in the region of maximum intensity, we observe that both radial and azimuthal polarization yield the same total rate of excitation. Furthermore, we can observe that the $W_{fi}^{LG}$ for cylindrical polarization is significantly more than circular polarization when the target is placed in the maximum intensity region of the beam with beam waist $w_{o}$ = 0.16~$\mu m$.

\section{Summary}\label{sec:summary}
We have theoretically investigated the photo-excitation of atoms by LG beams especially for cylindrical polarization. To do so, we constructed the complex weight factor of cylindrically polarized LG beam as a linear combination of complex weight factors for circular polarization. We analyzed strength of the electric-quadrupole field across the beam cross section of cylindrically polarized LG beams. We observed that the strength of the electric-quadrupole field of cylindrical polarization is maximum away from the beam axis. In addition, we observed that strength of the electric-quadrupole field in the beam cross section is sensitive to the radial index $p$ of the LG beam. The variation of the magnetic component of electric-quadrupole was analyzed as a function of beam waist $w_{o}$ and the radial distance $b$ from the beam axis. To better understand the variation of the magnetic component of electric-quadrupole field, we plotted the relative weight of the projection of electric-quadrupole $W_{r}(M)$ against the beam waist and the radial distance from the beam axis.
\par Furthermore, we illustrated the affects of varying multipole distribution on the magnetic sub-level population in the target atom. As an example, we considered the electric-quadrupole transition  in the target Ca$^{+}$ ion. In addition, we calculated total rate of excitation of electric quadrupole transition ($4s\; ^{2}S_{1/2} \rightarrow 3d\; ^{2}D_{5/2}$) in the target Ca$^{+}$~ion with a Gaussian spatial distribution. We observed that the total rate of excitation for the target Ca$^{+}$~ion excited by cylindrically polarized LG beam varies with beam waist $w_{o}$ and the distance between the center of the target and the beam axis $b_{o,x}$. Our calculation explicitly shows that the cylindrically polarized LG beams is more efficient than circular polarization for driving an electric quadrupole transition in a mesoscopic target.

\begin{acknowledgments}
This work has been funded by the Research School of Advanced Photon Science (RS-APS) of Helmholtz Institute Jena, Germany.

\end{acknowledgments}

\bibliography{Bibliography}% Produces the bibliography via BibTeX.

\end{document}